# The Viability of Life in Helium-Dominated Exoplanet Atmospheres

Sara Seager[a,b,c,*] and Janusz J. Petkowski[d,e]

[a] Department of Earth, Atmospheric and Planetary Sciences, Massachusetts Institute of Technology, 77 Massachusetts Avenue, Cambridge, MA 02139, United States
[b] Department of Physics, Massachusetts Institute of Technology, Cambridge, MA 02139, United States
[c] Department of Aeronautics and Astronautics, Massachusetts Institute of Technology, Cambridge, MA 02139, United States
[d] Faculty of Environmental Engineering, Wroclaw University of Science and Technology, 50-370, Wroclaw, Poland
[e] JJ Scientific, Mazowieckie, Warsaw 02-792, Poland

* Correspondence: seager@mit.edu

## Abstract

Helium is the second most abundant element in the Universe, yet helium-dominated atmospheres are rarely considered as environments for life. Two assumptions have contributed to this neglect: helium is expected to escape from rocky planets together with hydrogen, and Earth itself lacks a substantial helium envelope. Recent observations and atmospheric evolution models, indicate that rocky exoplanets can retain helium-dominated atmospheres. Here we make the case for life in such environments by synthesizing a largely overlooked body of laboratory research on life in helium-rich atmospheres. Every organism studied—including bacteria, fungi, protozoa, microalgae, plants, and animals—tolerates helium-dominated atmospheres when supplied with essential metabolic requirements. Across a range of pressures, helium has shown no demonstrated toxicity and no fundamental barrier to metabolism, cell division, photosynthesis, nitrogen fixation, or coordinated multicellular activity, including in humans. Helium-dominated atmospheres therefore constitute a viable and underrecognized planetary environment for life. Helium is chemically inert and spectroscopically unobtrusive, so the bulk atmosphere would neither contribute to the chemical destruction of biosignature gases nor strongly obscure their spectral features. Helium-rich atmospheres also have low mean molecular weights and large scale heights, strengthening spectral features in transmission and improving remote detectability.

**Highlights:**

- Helium-dominated atmospheres may provide habitable environments for life
- Helium is non-toxic to a wide range of known organisms, including humans
- Helium-rich atmospheres may favor the preservation and detection of biosignature gases
- Large atmospheric scale heights make helium-rich planets favorable targets for atmospheric characterization

## 1. Introduction

Every once in a while, a paradigm shift reveals a phenomenon that was present all along. We do not typically envision habitable planetary atmospheres dominated by helium. The conventional assumption holds that if light hydrogen gas escapes into space, any available inert helium should inevitably follow. Earth also offers little precedent, because our planet most likely has never possessed a substantial helium envelope.

Yet, the recent detection of escaping helium from the habitable-zone exoplanet LHS 1140b, interpreted as evidence for a helium-dominated upper atmosphere (Cherubim et al., 2026) presents a fundamental challenge to the existing assumptions. The milestone observation motivates us to not only scrutinize the physical formation pathways of helium-dominated atmospheres on exoplanets more closely, but also to evaluate whether a helium-dominated environment is fundamentally suitable for life.

Parallel to this astronomical revelation is an equally underappreciated, yet rich and focused, laboratory repository detailing microbial survival within pure helium or helium-dominated environments.

Are we undergoing a paradigm shift right now? We might be, though more empirical evidence is needed to know for certain. Here, we explore the question in three parts: we review the exoplanetary discoveries leading to the current shift in thinking about the existence of helium-dominated atmospheres (Section 2), we synthesize laboratory studies relevant to helium-dominated atmospheres (Section 3), and we conclude with a forward-looking discussion highlighting why helium is a favorable inert gas for detecting and sustaining a biosphere (Section 4).

## 2. Exoplanet Helium Atmospheres

The study of helium in exoplanet atmospheres and exospheres began with giant planets already expected to contain helium as their second most abundant constituent. (Seager and Sasselov, 2000) predicted the neutral-helium triplet at 1083 nm as an observable feature in transiting giant planets, specifically noting that helium could form an extended exosphere and produce exceptionally strong absorption.

If photoionized, helium atoms recombine to cascade down energy levels via both the singlet and triplet states. Because radiative decay from the metastable $2^3S$ state to the ground state is spin-forbidden, helium atoms accumulate in the $2^3S$ state. Stellar near-IR photons can then excite the $2^3S \rightarrow 2^3P$ transition, producing the 1083 nm absorption feature even if the helium gas itself is extremely tenuous.

Eighteen years later, (Spake et al., 2018) found a single spectral bin near 1083 nm with significantly enhanced absorption in a low-resolution Hubble transmission spectrum of the giant exoplanet WASP-107b. Subsequent high-resolution observations resolved the helium absorption signature, confirmed its planetary origin, and revealed blue-shifted absorption consistent with escaping helium; modeling indicated an extended thermosphere and a large exospheric comet-like tail (Allart et al., 2019). Later observations of WASP-107b traced helium absorption well beyond the end of the transit; modeling showed that the escaping helium forms a vast, comet-like tail extending to about seven planetary radii (Spake et al., 2021).

Since the initial (Spake et al., 2018) finding, observations of the 10830 nm triplet have detected helium absorption in more than twenty exoplanets (e.g., (Sanz-Forcada et al., 2025)), spanning hot Jupiters, warm Neptunes, and several sub-Neptune-sized exoplanets. Helium is now the leading observational tracer of atmospheric escape, providing constraints on outflow velocities, temperatures, mass-loss rates, and, in a few systems, enormous leading or trailing exospheric tails (e.g., (Allart et al., 2025)).

In parallel, atmospheric models demonstrated that helium-dominated exoplanet atmospheres can theoretically exist. The physical path begins with differential escape from a primordial hydrogen-helium envelope. When atmospheric loss approaches the diffusion-limited flux of hydrogen, the lighter hydrogen can escape more efficiently than helium. Helium then becomes progressively concentrated as the envelope evolves, eventually producing a helium-enriched or helium-dominated atmosphere if hydrogen removal continues long enough (Hu et al., 2015). This outcome requires escape to be strong enough to deplete hydrogen, but not so vigorous that helium is efficiently dragged away with it. Even 0.95–1.25 Earth-mass planets in the habitable zones of Sun-like stars may evolve through preferential hydrogen escape to form long-lived atmospheres with tens of bars of nearly pure helium (Lammer et al., 2025).

Not all planets expected to possess escaping helium exospheres show detectable helium absorption (see e.g., the recent survey by (Masson et al., 2024) and references therein). The helium absorption signal depends on many factors including: the stellar EUV and UV spectral shape, atmospheric helium abundance, mass-loss rate (Oklopčić and Hirata, 2018), and also stellar-wind interactions including instabilities and magnetic field interaction, temporal variability, and more (see e.g., (Hazra, 2025) and references therein.)

The recent detection of a time-variable escaping helium exosphere around the habitable-zone planet LHS 1140b provides evidence for a helium-dominated upper atmosphere on a predominantly rocky exoplanet (Cherubim et al., 2026). Models of the outflow favor an extremely hydrogen-depleted composition, with an H:He number ratio of order $10^{-3}$ and a mass-loss rate of order $10^{8}$ g s$^{-1}$. At this escape rate, the flow cannot entrain species heavier than about 9 atomic mass units, so carbon-, nitrogen-, and oxygen-bearing volatiles should remain at lower altitudes (Cherubim et al., 2026).

He-dominated exoplanet atmospheres may represent a fundamental planetary archetype rather than a rare evolutionary curiosity, based on recent population-level models coupling atmospheric escape to magma-ocean exchange (Cherubim et al., 2025). Specifically, the models predict a distinct concentration of He-rich planets near the upper edge of the small-planet radius valley, featuring median lifetimes of approximately 380

Myr around M dwarf stars (Cherubim et al., 2025). Under highly favorable conditions, more detailed evolutionary calculations demonstrate that preferential hydrogen loss can produce helium-dominated envelopes that persist for gigayear timescales (Hu et al., 2015). Interior sequestration of helium followed by helium exsolution may further enhance atmospheric helium at late evolutionary stages (Gupta et al., 2025).

Helium-dominated atmospheres may also be hiding among planets currently interpreted as highly metal-rich. Transmission spectra primarily constrain atmospheric scale height, so retrievals that fix the $He/H_2$ ratio near the solar value may compensate for an intermediate mean molecular weight by artificially increasing heavy-molecule abundances or invoking clouds. In a proof-of-concept retrieval of HD 209458b, allowing $He/H_2$ to vary produced He-rich solutions that required substantially less $H_2O$ than $H_2$-rich solutions to reproduce the same atmospheric scale height (de Wit et al., 2026). The degeneracy may be especially important for evolved Neptune- and sub-Neptune-sized planets such as GJ 436b, where preferential hydrogen escape offers a plausible route to helium dominance.

The astronomical observational evidence and modeling results therefore makes helium-dominated atmospheres physically plausible, but plausibility alone does not establish habitability. The next question is biological: does helium itself interfere with the processes required for life?

## 3. Laboratory Studies of Life in Helium-dominated Environments

We turn to the question of whether life can survive in helium-dominated atmospheres. The answer is unequivocally yes, based on a small but solid group of laboratory studies spanning decades. The initial motivation for older studies was for human exploration of the ocean and space: the physiology of deep-sea and saturation diving, in which $He-O_2$ is a standard breathing mixture (Doolette and Mitchell, 2011; Hess et al., 2006); the design of spacecraft cabin atmospheres (Hamilton Jr et al., 1971, 1970); and the algal oxygen-regeneration systems developed for space exploration (Ammann and Lynch, 1966; Myers, 1954) (see Section 3.5).

The studies collectively demonstrate that a broad range of cellular functions — respiration, photosynthesis, nitrogen fixation, cell division, and coordinated multicellular activity — can proceed in helium-dominated, and even pure helium, environments (see Table 1). Our extensive literature review shows that across every organism examined, no study identifies an inherently toxic effect or a metabolic barrier attributable to helium itself.

### 3.1 Prokaryotes Readily Survive and Grow in He-dominated Atmospheres

Laboratory evidence spanning more than half a century consistently demonstrates that helium-dominated atmospheres are fundamentally non-toxic to prokaryotes. *Escherichia coli*, cultivated under a 100% helium atmosphere at ambient pressure exhibits cellular generation times entirely comparable to those observed in regular air controls (Figure 1; (Seager et al., 2020)). While maximal cell densities in pure helium can be lower than in aerobic controls, the discrepancy represents a predictable energetic constraint of anaerobic metabolism rather than any physiological deficit or chemical inhibition imposed by the helium gas itself.

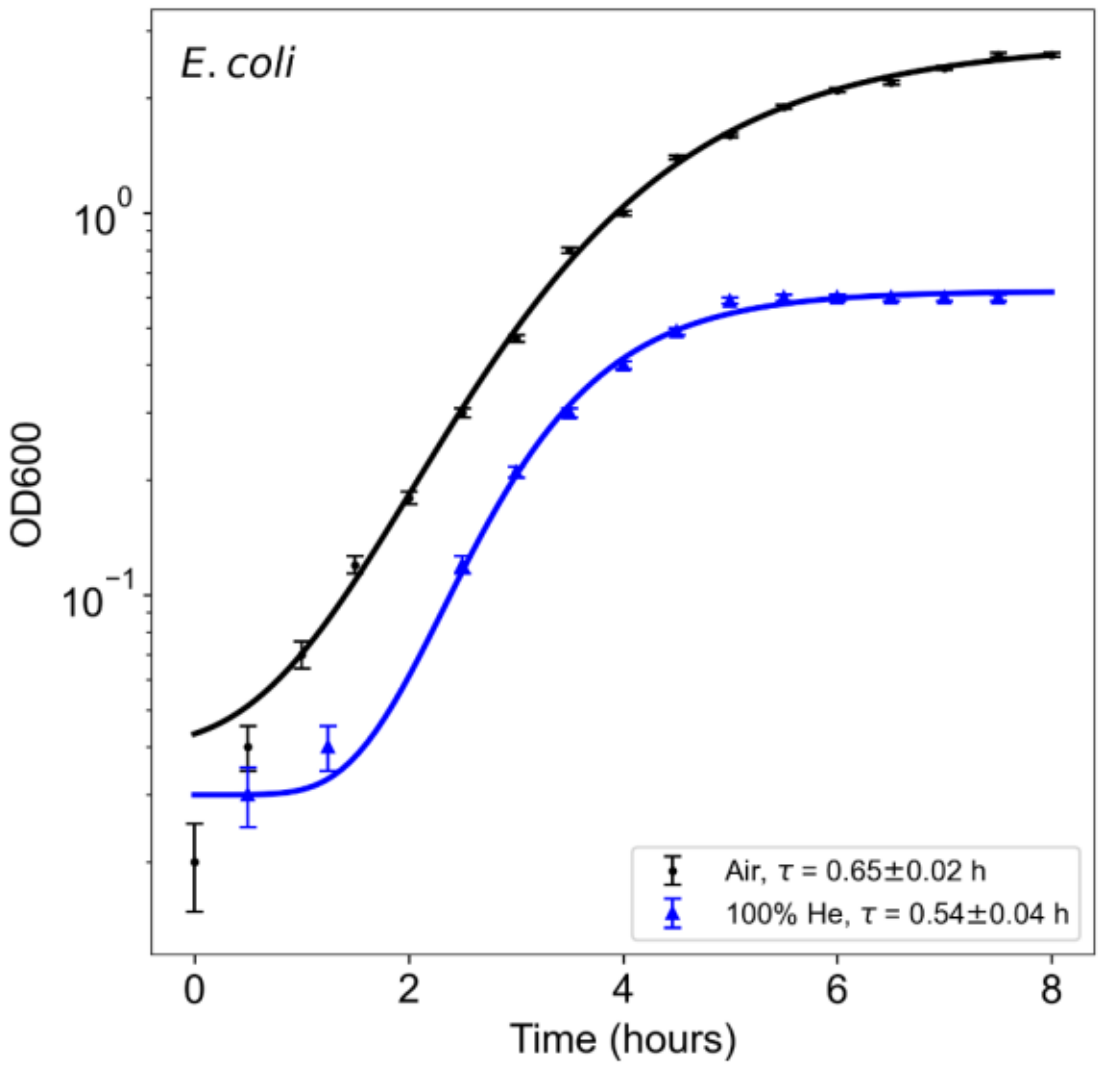

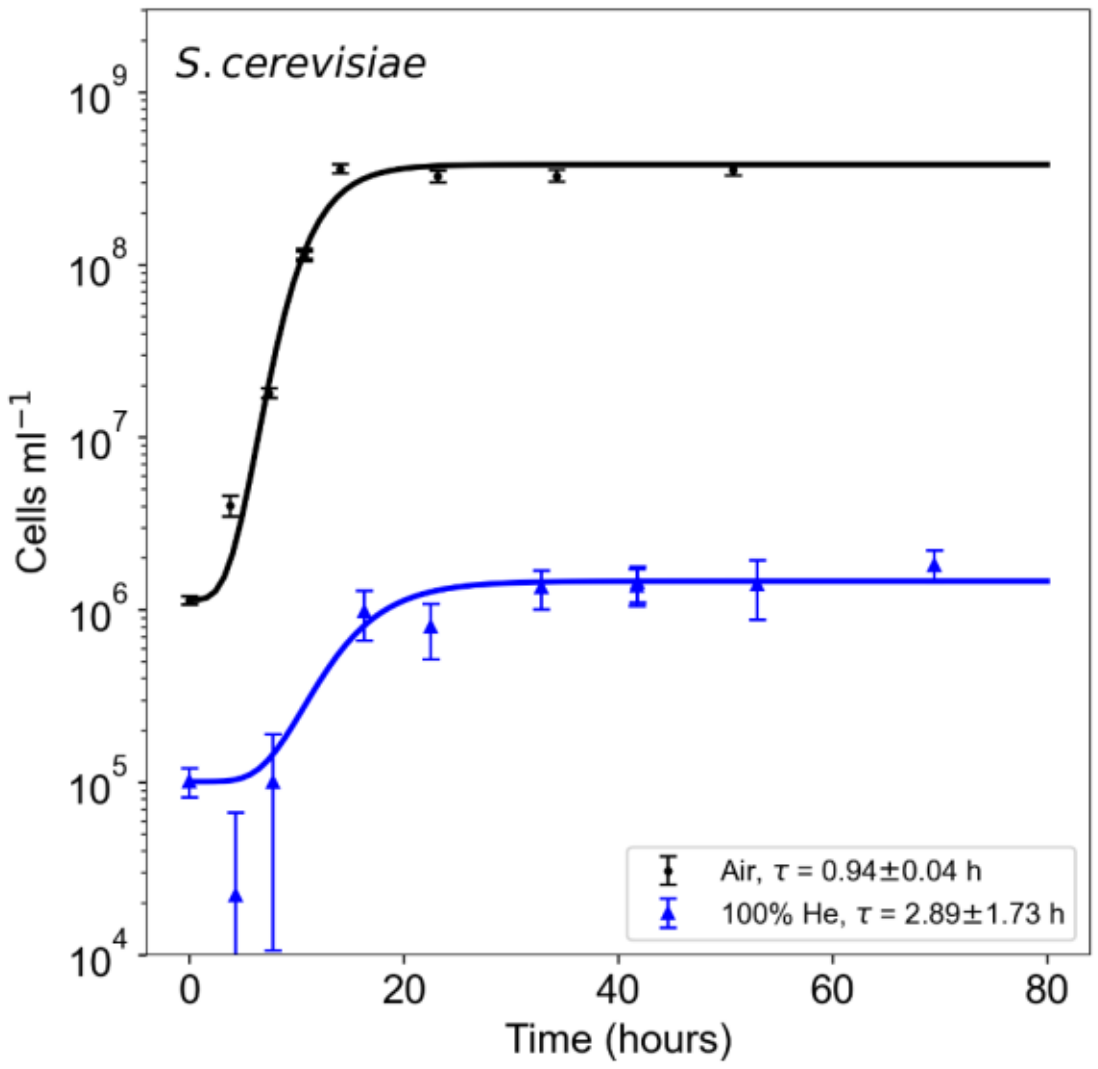

**Figure 1.** Growth of *E. coli* and *S. cerevisiae* (yeast) in a 100% He atmosphere with air as a control. Left panel: *E. coli* growth was measured by optical density at 600 nm (OD600), because the cells are too small for reliable direct counting with a haemocytometer. Error bars are reported at 2σ according to the machine measurement uncertainty. Right panel: yeast growth was calculated by manual cell counting and is reported as the number of cells per mL. The lower yeast cell counts in He reflect the absence of $O_2$ required for sterol and other biosynthetic pathways, rather than toxicity of helium itself. Curves show modified Gompertz fits to the measurements; error bars are 2σ uncertainties, and *τ* denotes the doubling time. Both organisms survived and reproduced in the He atmosphere. Data from (Seager et al., 2020).

Moreover, pure helium atmospheres at ambient pressures can also support critical biochemical pathways, such as nitrogen fixation. The cyanobacterium *Oscillatoria* sp. sustains nitrogenase activity under both 100% helium atmosphere and a He-$O_2$ mixture (Stal and Krumbein, 1985).

Prokaryote experiments consistently show that helium acts as a completely benign background gas that does not disrupt core biochemical machinery, even when compressed to dozens of atmospheres (Baden et al., 1981; Fenn and Marquis, 1968; MacNaughtan and Macdonald, 1982; Marquis et al., 1978; Schlamm et al., 1974; Taylor, 1979; Thom and Marquis, 1984)(see also Table 1).

Rather than functioning as a metabolic inhibitor, compressed helium frequently behaves as a neutral background medium that can even facilitate growth under physical stress. Spores of *Bacillus cereus* germinate unimpaired even when subjected to helium atmospheres at 100 atm pressure (Enfors and Molin, 1977). Pressures up to tens of atm prove to be entirely conducive for bacterial growth and non-mutagenic (Baden et al., 1981). Any physiological variations between ambient and high-pressure experiments strictly result from the high atmospheric pressure or localized anoxia and not from the helium gas itself. Helium is equally compatible with archaeal life. Deep-sea hyperthermophilic archaea, including the methanogen *Methanocaldococcus jannaschii* and *Sulfolobus acidocaldarius*, readily grow under high pressure helium atmospheres up to 750 atm (Miller et al., 1988; Nelson et al., 1992; Sturm et al., 1987).

A separate category of research focuses on He-$O_2$ mixtures, driven primarily by the operational needs of deep-sea and saturation diving where helium is substituted for nitrogen in an $N_2$-$O_2$ mixture to prevent narcosis at extreme depths (Doolette and Mitchell, 2011; Hess et al., 2006). Major findings indicate that helium is a physiologically benign carrier gas, with studies demonstrating that specialized bacteria can maintain metabolic activity under high-pressure He-$O_2$ environment, at 500 atm (Taylor, 1979).

### 3.2 Single-celled Eukaryotes Readily Grow in He-dominated Environments

Unicellular eukaryotes can survive and grow in pure helium atmospheres. Specifically, our own previous studies showed that the budding yeast *Saccharomyces cerevisiae* (a type of single-celled fungus) readily grows in a 100% helium atmosphere at ambient pressure (Seager et al., 2020). As expected, the growth rate and the maximal cell density of yeast cultivated in a 100% helium atmosphere at ambient pressure are lower when compared to growth values of yeast in air (Figure 1). Such a discrepancy is normal, and it is not an indication of He being toxic or less suited for life's survival and growth. We emphasize that $O_2$ is a critical substrate for the biosynthesis of many biochemicals required for yeast biology (e.g., sterols). In the absence of $O_2$ yeast have to rely on acquiring critical metabolites directly from the growth medium, importantly this limitation is unrelated to the role of $O_2$ in energy metabolism (Davies and Rine, 2006; Takishita et al., 2012; Waldbauer et al., 2011).

The only other study for yeast in a 100% helium atmosphere focused on high pressure (~60 atm), while including atmospheric pressure as a control. The study included both yeast and the single celled protozoan *Tetrahymena thermophila,* to directly demonstrate that helium does not disrupt eukaryotic cell division (Thom and Marquis, 1984). Research on another protozoan *Echinosphaerium nucleofilum*, however, showed that helium is not entirely inert — at high pressure (~80 atm) helium shortens the microtubule-based axopods, though He remains the least damaging of the inert gases tested (Miller et al., 1975).

Additional studies of unicellular eukaryotes in helium environments included a small amount of oxygen within the dominant helium environment. The filamentous fungus *Neurospora crassa* grows unimpaired in an atmosphere of approximately 95% helium with a few percent oxygen (Schreiner et al., 1962). Likewise, isolated cultures of human HeLa cells readily grow under He-$O_2$ atmosphere and experience reduced growth only at high pressures (~69 atm) (Bruemmer et al., 1967).

### 3.3 He-dominated Atmospheres Do Not Inhibit Photosynthetic Activity of Eukaryotic Microalgae

There are a surprisingly large number of studies that investigated helium tolerance of photosynthetic, oxygen-producing microorganisms.

Experiments on photosynthetic eukaryotes show that helium-dominated atmospheres do not inhibit photosynthetic activity. Eucaryotic microalgae readily survive and grow in He-dominated atmospheres. In aerobic experiments where He replaces $N_2$ in air, helium has no effect on the $O_2$ production, $CO_2$ consumption, or growth of microalgae such as *Chlorella* sp. (Ammann and Lynch, 1966; Orcutt et al., 1970). Unicellular microalgae, such as *Scenedesmus* sp. and *Chlamydomonas* sp. can also readily survive, photosynthesize and grow in 100% He atmosphere (Graves and Greenbaum, 1989; Lee et al., 1996; Zerveas et al., 2021).

Together, these experiments demonstrate that oxygenic photosynthesis and photoautotrophic growth — the metabolic foundation of a self-sustaining biosphere — proceed without impediment in helium-dominated atmospheres.

### 3.4 Multicellular Organisms Can Readily Survive in He-dominated Atmospheres

The tolerance of helium extends beyond single cells to complex multicellular life. Notably, all studies on the effects of helium on the survivability and physiology of complex life use He-$O_2$ gas mixtures due to reliance of Earth's multicellular life on atmospheric $O_2$. Although only a few studies exist, they all clearly show that He-dominated atmospheres do not have any toxic effects on multicellular life.

We showed that helium has no detrimental effects on adult fruit flies (*Drosophila melanogaster*) (Pajusalu et al., 2024). The flies survive helium-dominated, oxygen-poor atmospheres at ambient pressure, becoming immobile only under the most severe oxygen depletion (below approx. 33 mbar of $O_2$ (3.3%, the rest He)). The fruit flies recover movement as oxygen is restored; helium produces no apparent lasting detrimental effect on the flies beyond that of the accompanying hypoxia. We note that we chose fruit flies for these experiments as they are known to thrive, actively fly and breed at atmospheric $O_2$ concentrations as low as ~ 4% (the rest $N_2$) (Zhou et al., 2007). Our study goal was for active flight limits. Because the low mass density of a helium atmosphere reduces the aerodynamic force a beating wing can generate, sustained active flight requires a minimum atmospheric density (~0.93 kg $m^{-3}$ in these experiments).

Helium is also non-toxic to complex plants. In submerged aquatic plants, photosynthetic carbon dioxide uptake continues in a helium gas phase and is in fact enhanced relative to air, a consequence of the higher diffusivity of $CO_2$ in helium (Madsen and Maberly, 2003). We note however that the physical effects of helium are not always invariably benign to plants. Under prolonged anoxia, hydrated rye seeds (*Secale cereale*) appear to lose viability significantly faster in helium than in nitrogen (Latterell, 1966).

Experiments on animals also support the conclusion that helium is a non-toxic, inert gas. The early studies on excitable animal tissue prove that He-dominated atmosphere has no detrimental effects, showing that an isolated mammalian cardiac pacemaker continues to beat under He-$O_2$ atmosphere at pressures as high as 150 atm (Ornhagen, 1979).

Because the helium atom is light, helium conducts heat roughly five times more efficiently than $N_2$. A warm animal body therefore loses heat to a helium atmosphere considerably faster than to $N_2$-$O_2$ air. For the great majority of organisms, e.g., microbes, algae, plants, and ectothermic animals that simply equilibrate to ambient temperature, the efficient heat conductivity of helium is inconsequential (Cooper and Withers, 2026). For endotherms, such as mammals and birds, which expend energy to hold their body temperature above ambient, faster heat loss in the helium atmosphere raises the metabolic cost of thermoregulation and increases the metabolic rate (e.g., (Holloway and Geiser, 2001; Rosenmann and Morrison, 1974)). This additional metabolic cost however is readily met: rats and mice (including two successive generations of mice) raised for six months in a He-$O_2$ atmosphere at an ambient temperature grew at rates identical to mice kept in $N_2$-$O_2$ atmosphere, though the helium-exposed animals consumed more food and water to offset their greater heat loss (Hamilton et al., 1970). The thermal effect of helium-dominated atmosphere is therefore an easily manageable physiological effect on endothermic life, not a barrier to survival and growth.

### 3.5 Humans Live and Work for Weeks in He-dominated Atmospheres

Humans provide the clearest evidence that helium-dominated atmospheres are compatible with complex life. Indeed, as mentioned in earlier sections, the use of helium in deep-sea diving motivated several of the early studies of its effects on organisms. Commercial saturation divers routinely live and work for weeks at a time in pressurized habitats filled with He-$O_2$ ("heliox") mixtures (e.g., (Imbert et al., 2024, 2019; Łuczyński et al., 2019; Monnoyer et al., 2021; Thorsen et al., 1990)). Helium replaces the $N_2$ in the breathing atmosphere, while $O_2$ is maintained at a partial pressure sufficient to support normal respiration. In saturation diving, workers repeatedly travel from the pressurized living chamber to work underwater at depth and then return without decompressing to surface pressure after each excursion; helium replaces nitrogen because nitrogen is narcotic at high pressure and increases the risk of decompression sickness ("the bends"). Divers therefore remain continuously in the helium-dominated atmosphere, where they eat, sleep, exercise and work, for periods of up to 28 days before a single final decompression to surface pressure.

The physiological challenges of saturation diving arise primarily from high pressure, elevated oxygen exposure, decompression, confinement and the demanding working environment, rather than from helium toxicity itself. Because helium conducts heat much more efficiently than $N_2$ (see section 3.4), saturation habitats must be kept unusually warm, typically around 30-34 °C (86-93 °F), to limit body heat loss.

The routine use of saturation diving with heliox therefore provides decades of direct human experience with prolonged helium-dominated atmospheres. Humans demonstrate at the organismal level what the microbial, plant and animal experiments described above demonstrate experimentally: replacing the dominant inert component of an atmosphere with helium does not intrinsically prevent survival or normal biological function.

## 4. Discussion

### 4.1 Helium Effects May be Positive for Biology

So far in our review of the effects of helium on living organisms, we have focused on helium gas as chemically inert, non-toxic, and otherwise biologically unobtrusive. Helium's high diffusivity, however, can have a significant effect on the physiology of plants and animals. Early studies discovered, rather unexpectedly, that chicken eggs incubated in a He-$O_2$ atmosphere hatch at only about 45% the rate of $N_2$-$O_2$ air controls (Weiss et al., 1965; Weiss and Wright, 1968). Notably, chicks that do hatch grow normally in the He-$O_2$ atmosphere (Rhoades et al., 1967). The reason for the hatching failure can be solely attributed to the much higher diffusivity of helium, as compared to nitrogen (Erasmus and Rahn, 1976). Eggs incubated in helium lose water to diffusion at roughly twice the rate of eggs incubated in $N_2$-$O_2$ atmosphere (Weiss and Wright, 1968). This apparent detrimental effect of He-$O_2$ atmosphere on chicken embryonic development can be readily mitigated by decreasing the diffusion of gases through eggshells. Covering approximately 50% of the eggshell surface with melted paraffin improves the hatch rate to $N_2$-$O_2$ control values (Weiss, 1975). These studies further support the conclusion that He gas is biochemically inert, and that any effects that helium has alter the physical conditions of the environment, not the chemistry.

We emphasize however that the higher diffusivity of gases through helium may have also beneficial effects on life when helium is the dominant atmospheric gas (Madsen and Maberly, 2003). The photosynthetic $CO_2$ uptake in some submerged aquatic plants is enhanced relative to air, which (Madsen and Maberly, 2003) attribute to the higher diffusivity of $CO_2$ through helium. Because the helium atom is light, gases diffuse faster through a helium-dominated atmosphere than through air, so any process rate-limited by the transport of a gas to or from a surface should accelerate accordingly. We therefore regard helium's high diffusivity as a genuine advantage by which a helium-dominated atmosphere could favor, rather than merely tolerate, life.

### 4.2 Atmosphere Characteristics for Habitability and Observation

The inertness of helium is a key factor for habitability of planets with helium-dominated atmospheres. Helium is chemically inert and so helium functions purely as a background or "filler" gas, much as molecular nitrogen ($N_2$) does in Earth's atmosphere. Helium is also spectroscopically unobtrusive, so the bulk atmosphere would neither contribute to the chemical destruction of biosignature gases nor strongly obscure their spectral features.

We emphasize, however, that helium is significantly more chemically inert than $N_2$. Helium, as a monoatomic gas, cannot be destroyed by lightning and other high-energy processes. Helium neither reacts with other atmospheric gases nor undergoes photodissociation, so helium is not depleted by atmospheric chemistry. Helium can coexist with virtually any other atmospheric gas, including $N_2$, $CO_2$ and even reactive gases, like ozone ($O_3$) and molecular oxygen.

A He-$O_2$ atmosphere analogous to Earth's $N_2$-$O_2$ atmosphere is therefore chemically stable and fully compatible with aerobic metabolism and with oxygenic photosynthesis (see Table 1). In other words, from a chemical point of view helium-dominated atmosphere is as compatible with life as $N_2$-dominated atmospheres.

Helium is a poor greenhouse absorber, which sharply distinguishes He-dominated atmospheres from $H_2$-dominated atmospheres. $H_2$-$H_2$ collision-induced absorption (CIA) provides substantial infrared opacity and can act as a strong, noncondensing greenhouse in atmospheres with high enough pressure, allowing sufficiently massive $H_2$ atmospheres to maintain surface liquid water far beyond the classical habitable zone (Pierrehumbert and Gaidos, 2011). Helium lacks this strong $H_2$-$H_2$ CIA warming, so maintaining habitable surface temperatures on a helium world depends largely on radiatively active gases such as $H_2O$, $CO_2$ and $CH_4$, rather than on He itself. One interesting nuance is that a He-dominated atmosphere can nevertheless be modestly warmer than an otherwise comparable $N_2$-dominated atmosphere because monatomic He has a steeper tropospheric adiabat, with weaker Rayleigh scattering also contributing. In the 1-bar models of (Kuss et al., 2026), the effect is only of order 10-20 K in surface temperature, depending on the stellar spectrum.

Helium has another potentially important property for planetary habitability: it remains gaseous under all plausible surface conditions on a habitable planet. A helium-dominated atmosphere therefore cannot undergo the condensation-driven collapse possible for atmospheres whose dominant constituent can freeze or condense. This may be particularly important for synchronously rotating planets around M dwarfs, where permanent nightsides can become extremely cold when day-to-night heat transport is inefficient (Joshi et al., 1997; Wordsworth, 2015). A helium-dominated atmosphere would retain its bulk gaseous envelope even under such conditions. This does not, however, guarantee a habitable climate, because $H_2O$, $CO_2$ and other condensable species could still become cold-trapped on the nightside if the atmosphere were too thin or radiatively weak.

Beyond avoiding atmospheric collapse, helium's thermodynamic properties may also help limit nightside cooling. Helium has a mass-specific heat capacity approximately five times larger than $N_2$, which can substantially lengthen the atmospheric radiative cooling timescale and may influence day-to-night heat redistribution. A useful comparison is between the characteristic radiative cooling and advective timescales,
$\tau_{\mathrm{rad}} \approx \frac{c_p\, p}{4\, g\, \sigma\, T^3}$ and $\tau_{\mathrm{adv}} \approx \frac{L}{U}$, where ($p$) is atmospheric pressure, ($c_p$) is the mass-specific heat capacity, ($g$) is surface gravity, ($T$) is atmospheric temperature, ($L$) is the characteristic transport distance, and ($U$) is the characteristic wind speed. At the same pressure and temperature, the larger heat capacity of helium compared to $N_2$ substantially increases its radiative cooling timescale. Whether this translates into more efficient day-to-night heat redistribution is less straightforward, because the advective velocity depends on the atmospheric circulation itself. The resulting day-night temperature contrast also depends strongly on infrared opacity, surface pressure and surface–atmosphere coupling (e.g., (Koll and Abbot, 2016)). Dedicated climate modeling is therefore needed to determine whether He-dominated atmospheres are systematically more effective than $N_2$-dominated atmospheres at maintaining warm nightsides on synchronously rotating planets.

Helium-dominated atmospheres have a vertically extended atmosphere due to Helium's low mean molecular weight. The atmospheric scale height is inversely proportional to the atmospheric mean molecular weight, so at the same temperature and surface gravity, replacing an $N_2$-dominated atmosphere with helium increases the scale height by about a factor of seven. Transmission features from trace gases would therefore be substantially larger than in an otherwise comparable $N_2$-rich atmosphere, improving the detectability of minor atmospheric constituents and potential biosignature gases.

### 4.3 Summary

The possibility of helium-dominated, temperate exoplanet atmospheres asks us to widen a familiar assumption: that life-bearing atmospheres must resemble Earth's in their bulk composition. Helium contributes little chemistry of its own, yet that very inertness allows the gases required by life to coexist within a stable atmospheric envelope, while its low molecular weight may make their spectral signatures unusually accessible.

The fact that such diverse forms of life can survive, metabolize, and reproduce under helium-dominated conditions supports the inclusion of helium-dominated exoplanets among the targets in the search for life beyond Earth.

The convergence of atmospheric-evolution models, the first observations of helium around a habitable-zone rocky planet, and decades of laboratory evidence establishes that helium is not an exotic obstacle to life and might in fact enable the first biosignature gas detection on a rocky world in our Galaxy.

**Table 1.** Summary of Laboratory Studies on the Viability of Life in Helium-dominated Atmospheres.

| Organism | Type | He Level | $O_2$ Status | Pressure | Endpoint Measured | Effect of He and Outcome | Reference |
|---|---|---|---|---|---|---|---|
| **Helium-dominant atmospheres at near-ambient pressure (direct analogues)**<br>*Helium is the bulk atmospheric gas (≥ ~78% He) at ~1 atm. Closest laboratory analogues to a temperate He-dominated atmosphere.* | | | | | | | |
| *Escherichia coli* | Prokaryote (bacteria) | 100% | Anoxic (trace $O_2$) | ~1 atm | Growth ($OD_{600}$) | Growth rate in 100% He comparable to air ($N_2$-$O_2$ mixture); lower maximum cell density (anoxic). | (Seager et al., 2020) |
| *Saccharomyces cerevisiae* | Eukaryote (yeast) | 100% | Anoxic (trace $O_2$) | ~1 atm | Growth (manual cell count) | Growth rate in 100% He slower than in air ($N_2$-$O_2$ mixture); lower maximum cell density (anoxic). | (Seager et al., 2020) |
| *Saccharomyces cerevisiae* | Eukaryote (yeast) | 80% | Aerobic (20% $O_2$) | ~1 atm | Survival and $O_2$ consumption | He effectively inert. No change in $O_2$ consumption. | (Cook, 1950; Maio and Neville, 1967) |

| Organism | Type | He Level | $O_2$ Status | Pressure | Endpoint Measured | Effect of He and Outcome | Reference |
|---|---|---|---|---|---|---|---|
| *Chlorella pyrenoidosa* | Eukaryote (microalga) | ~78% | Aerobic (~20% $O_2$, 2% $CO_2$) | ~1 atm | Growth + $O_2/CO_2$ exchange | No effect of He on $O_2$ production, $CO_2$ consumption, or growth over 3 weeks. | (Ammann and Lynch, 1966) |
| *Chlorella sorokiniana* | Eukaryote (microalga) | ~74-98% | Aerobic (+$O_2$), Anaerobic (+$CO_2$) | ~0.25-3 atm | Growth | He effectively inert. | (Orcutt et al., 1970) |
| *Scenedesmus obliquus* | Eukaryote (microalga) | 100% | Anoxic | ~1 atm | Growth + photosynthesis | Viable, photosynthetically active; Growth in He slower than in air ($N_2$-$O_2$ mixture). | (Zerveas et al., 2021) |
| *Chlamydomonas reinhardtii* | Eukaryote (microalga) | ~99.9% | Anaerobic (+700 ppm $CO_2$) | ~1 atm | Photoautotrophic growth | Photoautotrophic growth in He comparable to air ($N_2$-$O_2$ mixture). | (Lee et al., 1996) |
| *Neurospora crassa* | Eukaryote (fungus) | ~95% | Aerobic (~5% $O_2$) | ~1 atm | Growth (colony extension) | Growth rate scaled inversely with inert-gas mass — fastest in helium of all gases tested. | (Schreiner et al., 1962) |
| *Chlamydomonas reinhardtii* | Eukaryote (microalga) | ~100% carrier | Anoxic (+300 ppm $CO_2$) | ~1 atm | Photosynthesis ($O_2$ evolution) | Activity only: photosynthetic $O_2$ evolution detected in a He atmosphere. | (Graves and Greenbaum, 1989) |
| *Oscillatoria sp.* | Prokaryote (cyanobacterium) | 100% / 80% | ± 20% $O_2$ | ~1 atm | Nitrogenase ($N_2$ fixation) | Activity only: nitrogenase (acetylene reduction) functional under He and He/$O_2$. | (Stal and Krumbein, 1985) |
| Submerged aquatic macrophytes | Eukaryote (plant, multicellular) | ~100% carrier | Variable $CO_2$ (vs air, $N_2$) | ~1 atm | $CO_2$ uptake (gas exchange) | Activity only: photosynthetic $CO_2$ uptake functional; He raised $CO_2$ uptake (faster $CO_2$ diffusion) | (Madsen and Maberly, 2003) |
| *Drosophila melanogaster* | Animal (insect, multicellular) | ~100% → air | Trace → 21% (continuum) | ~1 atm | Survival and active flight | Survived; immobilized at lowest $O_2$, recovered as $O_2$ returned. He non-toxic. | (Pajusalu et al., 2024) |
| *Drosophila melanogaster, Zootermopsis nevadensis,* | Animal (insect, multicellular) | 80% | Aerobic (20% $O_2$) | ~1 atm | Survival and $O_2$ consumption | He effectively inert. No change in $O_2$ consumption. Decrease in total development time in *D. melanogaster.* | (Cook, 1950) |
| *Tenebrio molitor* | Animal (insect, multicellular) | 80% | Aerobic (20% $O_2$) | ~1 atm | Survival and $O_2$ consumption | $O_2$ consumption accelerated. $CO_2$ production rate dependent on the life cycle stage. Decrease in total development time. | (Cook, 1950) |
| *Coleonyx variegatus, Cnemidophorus tesselatus* | Animal (reptile, multicellular) | 80% | Aerobic (20% $O_2$) | ~1 atm | Survival and $O_2$ consumption | He effectively inert. No change in $O_2$ consumption. Decrease in $CO_2$ output. | (Cook, 1950) |
| *Gallus gallus* | Animal (bird, multicellular) | 80% | Aerobic (20% $O_2$) | ~1 atm | Survival and metabolic rate | Higher metabolic rate than in air ($N_2$-$O_2$ mixture). | (Rhoades et al., 1967) |
| Cultured mammalian cells; mouse tissue slices | Animal cells/tissue (in vitro) | 80% (variable He) | Aerobic (variable $O_2$) | ~1 atm to high pressure | Proliferation; $O_2$ consumption / glycolysis | $O_2$ consumption accelerated. | (Cook and South JR, 1953; Leon and Cook, 1960; Maio and Neville, 1967; South Jr and Cook, 1954; Stephenson, 1969) |
| *Rattus* sp., *Mus* sp. (rats, mice) | Animal (mammal, multicellular) | 80% | Aerobic (20% $O_2$) | ~1 atm | Growth, reproduction | Growth rates and reproduction are identical to air ($N_2$-$O_2$ mixture); no biochemical change; increased food/water intake (higher heat loss). | (Hamilton et al., 1970; Rhoades et al., 1967) |
| **High-pressure inert-gas studies (helium atmosphere tested at elevated total pressure)** *Helium tested at high total pressure. Collectively the results show that helium is biologically inert.* | | | | | | | |
| *Escherichia coli* | Prokaryote (bacteria) | ~99.7% | 0.2 atm. $O_2$ | 68 atm | Growth | Normal growth (growth lag shortened 1-1.5 h; growth rate/yield unchanged as compared to air ($N_2$-$O_2$ mixture). | (Schlamm et al., 1974) |
| *Bacillus cereus* | Prokaryote (bacteria) | 100% | Anoxic | 100 atm | Germination | He effectively inert. | (Enfors and Molin, 1977) |
| *S. faecalis, E. coli, S. aureus* | Prokaryote (bacteria) | ~95–99% (+air) | Aerobic (air) | 20-70 atm | Growth | He stimulated growth (faster exponential phase); He potentiated $O_2/N_2O$ inhibition. | (Marquis et al., 1978) |

| Organism | Type | He Level | $O_2$ Status | Pressure | Endpoint Measured | Effect of He and Outcome | Reference |
|---|---|---|---|---|---|---|---|
| *Streptococcus faecalis* | Prokaryote (bacteria) | Pressurizing gas | Aerobic | ≤41 atm | Growth | He effectively inert. | (Fenn and Marquis, 1968) |
| *E. coli, S. cerevisiae, Tetrahymena thermophila* | Prokaryote + eukaryotes | ~100% | Aerobic culture | ≤~60 atm | Growth | He effectively inert. | (Thom and Marquis, 1984) |
| *Salmonella typhimurium* | Prokaryote (bacteria) | ~98–99% (+air) | Aerobic (air) | 50-100 atm | Growth + mutagenesis | He effectively inert. No mutagenicity, no viability loss. Normal growth at 50 atm., decreased growth at 100 atm. | (Baden et al., 1981) |
| Marine bacterium "EP-4" | Prokaryote (bacteria) | ~99% (He-$O_2$) | Controlled low $O_2$ | 500 atm | Growth | Reduced growth at high pressures. | (Taylor, 1979) |
| *Acholeplasma laidlawii* | Prokaryote (bacteria) | Pressurizing gas | Aerobic culture | 300 atm | Growth | Reduced growth at high pressures. | (MacNaughtan and Macdonald, 1982) |
| *Methanocaldococcus jannaschii,* | Prokaryote (archaea) | Pressurizing gas | Anaerobic (+$H_2$/$CO_2$ substrate) | up to ~750 atm | Growth, methanogenesis | Growth and methanogenesis accelerated by hyperbaric He (up to ~750 atm); He the least-inhibitory pressurizing medium. | (Miller et al., 1988) |
| *Sulfolobus acidocaldarius,* | Prokaryote (archaea) | Pressurizing gas | Aerobic culture | up to ~120 atm | Growth | He effectively inert. | (Sturm et al., 1987) |
| Hyperthermophilic archaeon “ES4” | Prokaryote (archaea) | Pressurizing gas | Anaerobic | 500 atm | Growth | Reduced growth at high pressures. High-pressure He raised gas production and the maximum growth temperature. | (Nelson et al., 1992, 1991) |
| *Neurospora crassa* | Eukaryote (fungus) | Pressurizing gas | Aerobic | up to ~300 atm | Growth | Reduced growth at high pressures. | (Buchheit et al., 1966) |
| *Tetrahymena pyriformis* | Eukaryote (ciliate) | Pressurizing gas | Aerobic culture | ≥175 atm | Cell division rate | Reduced growth at high pressures. | (Macdonald, 1975) |
| *Paramecium multimicronucleatum* | Eukaryote (ciliate) | Pressurizing gas | Aerobic culture | 52 atm | Cell movement | He effectively inert. Cell movement unaffected. | (Sears and Gittleson, 1964) |
| *Spirostomum ambiguum* | Eukaryote (ciliate) | Pressurizing gas | Aerobic culture | 10-120 atm | Cell movement | He effectively inert. Cell movement reduced at high pressures. | (Macdonald and Kitching, 1976) |
| *Echinosphaerium nucleofilum* | Eukaryote (heliozoa) | Pressurizing gas | Aerobic culture | 10-130 atm | Axopod length; cell integrity | Helium shortens the microtubule-based axopods; helium is the least damaging of the inert gases. | (Miller et al., 1975) |
| Human HeLa cells | Eukaryote (human) | 75% | Aerobic (~20% $O_2$, 5% $CO_2$) | 69 atm | Cell division rate | Reduced growth at high pressures. | (Bruemmer et al., 1967) |
| *Arbacia punctulata* | Eukaryote (sea urchin) | Pressurizing gas | Aerobic culture | 61 atm | Cell division | Normal embryo cell division. | (Haywood, 1953) |
| Mouse sinus-node (pacemaker) tissue | Animal tissue (ex vivo) | Pressurizing gas | Perfusate 98% $O_2$/2% $CO_2$ | up to ~150 atm | Cardiac beating frequency | Activity only: excised tissue kept beating. | (Ornhagen, 1979) |

## Data availability statement

All data is included in the article.

## CRediT authorship contribution statement

**Sara Seager**: Conceptualization, Writing - Original Draft, Writing - Review & Editing.
**Janusz J. Petkowski**: Writing - Original Draft, Writing - Review & Editing.

## Declaration of Competing Interest

The authors declare that they have no known competing financial interests or personal relationships that could have appeared to influence the work reported in this paper.

## Acknowledgements
We thank Collin Cherubim for useful discussions.